\documentclass[12pt,a4paper]{article}
\usepackage{graphicx,subfigure,amsmath}
\usepackage{geometry}
\usepackage{feynmp-auto}
\usepackage{hyperref}
\usepackage{authblk}

\begin{document}
\unitlength=1mm
\title{\textbf{A no-go theorem for the dark matter interpretation of the positron anomaly}}
\author[1,2]{Maxim Laletin}
\affil[1]{Space sciences, Technologies and Astrophysics Research (STAR) Institute, Universit\'{e} de Li\`{e}ge, B\^{a}t B5A, Sart Tilman, 4000 Li\`{e}ge, Belgium}
\affil[2]{National Research Nuclear University MEPhI, 115409 Kashirskoe shosse 31, Moscow, Russia}
\date{}

\maketitle

\begin{abstract}

The overabundance of high-energy cosmic positrons, observed by PAMELA and AMS-02, can be considered as the consequence of dark matter decays or annihilations. We show that recent FERMI/LAT measurements of the isotropic diffuse gamma-ray background impose severe constraints on dark matter explanations and make them practically inconsistent. %A possible way-out for dark matter explanation is related to the hypothesis of a dark disk. 
\end{abstract}
\baselineskip=14pt
\vspace{10pt}

\section{Introduction}

\begin{figure}[h!]\label{FSRdiag}
\centering
\begin{fmffile}{diagramFSR}
\fmfframe(5,5)(5,5){
\begin{fmfgraph*}(70,40)
% Note that the size is given in normal parentheses
% instead of curly brackets.
% Define external vertices from bottom to top
\fmfpen{thin}
\fmfleft{i1,i2}
\fmflabel{DM}{i1}
\fmflabel{DM}{i2}
\fmfright{o1,o2,o3,o4}
\fmflabel{$X$}{o1}
\fmflabel{$e^+$}{o2}
\fmflabel{$\nu_e$}{o4}
\fmflabel{$\gamma$}{o3}
\fmf{fermion}{i1,v2,i2}
%\fmf{photon}{v1,v2}
\fmfblob{.15w}{v2}
\fmf{photon,tension=0.5,label=$W^+$,side=up}{v2,v3}
\fmf{fermion}{o2,v4,v3,o4}
%\fmf{fermion}{v2,o3}
\fmffreeze
\fmf{photon}{v4,o3}
\fmf{fermion}{v2,o1}
%\fmf{fermion}{v2,o5}
\fmfi{plain}{vpath (__v2,__o1) shifted (thick*(0,2))}
\fmfi{plain}{vpath (__v2,__o1) shifted (thick*(-1,-2))}
\end{fmfgraph*}
}
\end{fmffile}
\caption{A diagram illustrating an example of the DM annihilation process providing a positron via $W^+$ decay and some variety of states $X$. The positron emits final state radiation (FSR).}
\end{figure}
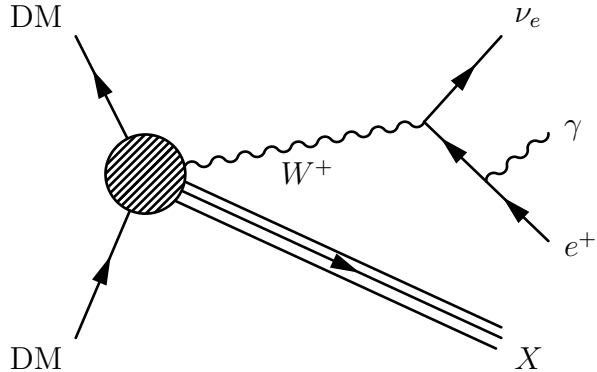

\nocite{Belotsky:2016tja}
The unexpected increase of the positron fraction in cosmic rays with energies above 10 GeV (also known as the ``positron anomaly'') was observed for the first time in the PAMELA experiment \cite{Adriani:2008zr} and was later confirmed by AMS-02 \cite{Aguilar:2013qda}. A lot of attention was paid to this discovery since the standard mechanisms of positron production and acceleration predicted a much steeper energy spectrum of cosmic positrons. The list of possible explanations includes, inter alia, decays or annihilations of dark matter (DM) particles, implying the existence of interconnection between our world and ``dark world''. This intriguing possibility is though highly constrained by a set of direct, indirect and accelerator-based observations, which force DM models to become more and more sophisticated. But no matter how complicated a DM model explaining the positron anomaly is, it should obviously fulfill the principal requirement that it produces a sufficient amount of high-energy positrons. The undesirable consequence of this fact is that, regardless of the prior (internal) processes, production of charged particles is accompanied by gamma-ray emission (see Fig.~\ref{FSRdiag}).

In addition, gamma rays are produced during the propagation of charged particles through the Galactic gas and the electromagnetic media, mainly in such processes as Bremsstrahlung and inverse Compton scattering (ICS). As we are going to show, even this at first sight small contribution to Galactic gamma rays may come in conflict with the latest Fermi-LAT data on the isotropic diffuse gamma-ray background \cite{Ackermann:2014usa,DiMauro:2016cbj} and, furthermore, rule out DM explanations of the high-energy cosmic positron excess. Basically, the reason of this problem is the following: %the rate of gamma rays production is proportional to the rate of positron production and the latter is ``fixed'' by cosmic positron measurements. However, 
the total amount of positrons and photons %also 
depends on the size of the volume in which their sources are concentrated, and though physically both positrons and photons have the same source, %their ``effective'' volume (meaning 
the volume of space from which they mostly arrive is substantially different. While only those positrons that were produced in the $\sim 3$ kpc proximity can approach the Earth (due to their stochastic motion in the Galactic magnetic fields and the corresponding energy losses), gamma rays can come to us directly from any point of the DM halo, where they were born. Now, since the DM halo is indeed large, the amount of gamma rays can simply overwhelm the observed limits. %As mentioned above, the emergence of these gamma rays is an ``irremovable'' fact.

\section{The theorem}

The no-go theorem we are considering can be expressed as follows:
\vspace{\baselineskip}

\textit{Any model of DM %(regardless of the particle variety or interaction properties)
	 providing a satisfactory explanation of the high-energy cosmic positron data and assuming an isotropic distribution of annihilating or decaying DM particles in the Galactic halo produces an overabundance of gamma rays that contradicts the latest experimental data on the diffuse gamma-ray background.}

\begin{figure}[t]
%\centering
%\vspace{0.2 cm}
%\hspace{0.5 cm}
\begin{minipage}[h]{0.49\linewidth}
\center{\includegraphics[width=1\textwidth]{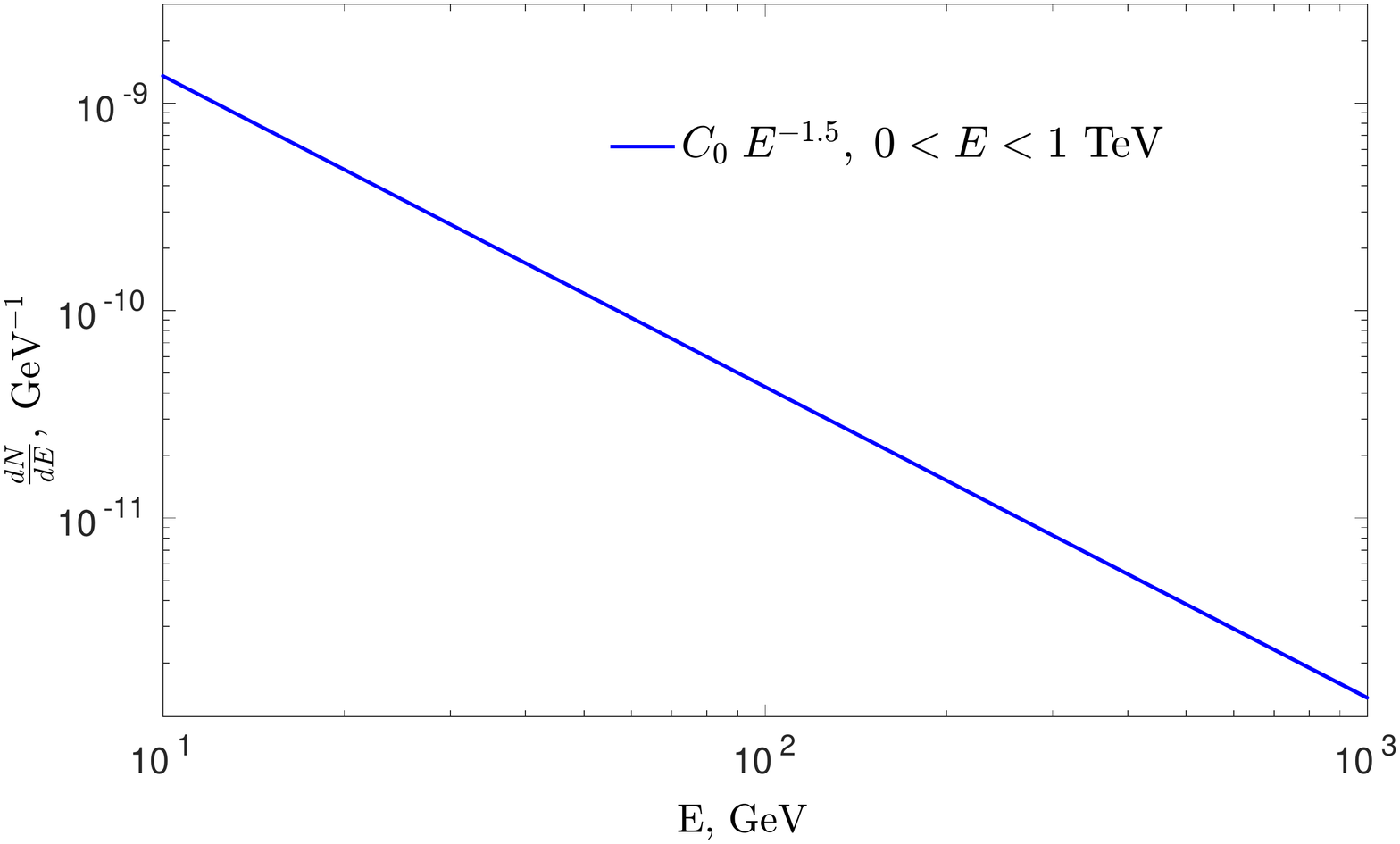}}
\end{minipage}
%\hfill
\begin{minipage}[h]{0.49\linewidth}
\centering{\includegraphics[width=1\textwidth]{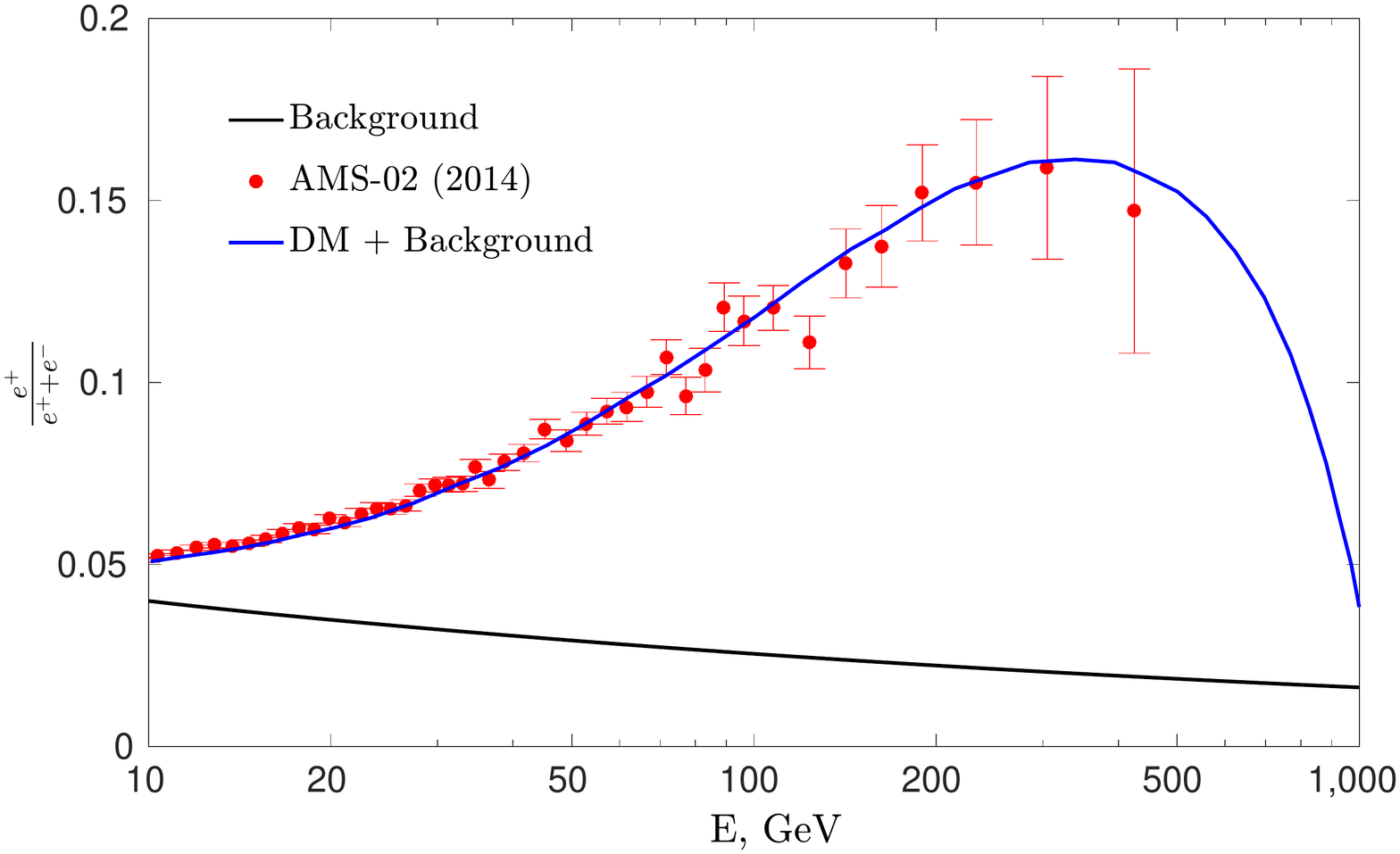}}
\end{minipage}
\caption{\label{minpos}  
DM positron injection spectra (left), providing the best possible fit to the AMS-02 positron fraction data (right).}
\end{figure}

\vspace{\baselineskip}
\noindent The proof starts with ``the extraction'' of the initial (injection) spectrum of positrons produced in DM annihilations  from the cosmic positron data (decays result in larger values of gamma-positron ratio compared to the case of annihilations and hence are discarded right away). Though, technically we did it the other way round (see \cite{Belotsky:2016tja} for the details) -- we found the injection spectrum, which eventually (after taking into account the effects of propagation) provides the best possible fit to the AMS-02 data on cosmic positron fraction (Fig.~\ref{minpos}). 

To calculate the local fluxes of positrons (and the ICS and Bremsstrahlung contributions to gamma rays) from DM annihilations, the GALPROP code was used \cite{GalpropCode}. One may argue that our results depend on the choice of propagation parameters (spatial diffusion coefficient, size of magnetic halo, etc.), which are not really well defined yet (we used the set of propagation parameters providing the best fit of AMS-02 proton and Boron-to-Carbon data \cite{Jin:2014ica}). We agree that a piece of uncertainty comes from the propagation model, though we do not expect it to influence the result significantly. In other words, a ``finely tuned'' propagation model is not likely to solve the problem with gamma rays. 

\begin{figure}[h!]
	\centering
	\begin{fmffile}{Vertex1}
		\fmfframe(5,5)(5,5){
			\begin{fmfgraph*}(40,30)
				% Note that the size is given in normal parentheses
				% instead of curly brackets.
				% Define external vertices from bottom to top
				%\fmfpen{thin}
				\fmfleft{i1}
				\fmfright{o1,o2,o3}
			    \fmflabel{$\phi$}{i1}
			    \fmflabel{$e^+$}{o3}
				\fmflabel{$\gamma$}{o2}
				\fmflabel{$f$}{o1}	
				\fmf{scalar,straight}{i1,v1}	   
				\fmfdot{v1}
				\fmf{fermion}{o1,v2}
				\fmf{plain}{v2,v1}
				\fmf{fermion}{v1,v3,o3}
				\fmffreeze
				\fmf{photon}{v3,o2}
				%\fmfi{boson}{vpath (__i1,__v1) shifted (thick*(0.5,2))}
			\end{fmfgraph*}
		}
	\end{fmffile}
	\hspace{15pt}
	\begin{fmffile}{Vertex2}
		\fmfframe(5,5)(5,5){
			\begin{fmfgraph*}(40,30)
				% Note that the size is given in normal parentheses
				% instead of curly brackets.
				% Define external vertices from bottom to top
				%\fmfpen{thin}
				\fmfleft{i1}
				\fmfright{o1,o2,o3}
				\fmflabel{$f$}{i1}
				\fmflabel{$e^+$}{o3}
				\fmflabel{$\gamma$}{o2}
				\fmflabel{$\phi$}{o1}	
				\fmf{fermion}{i1,v1}	   
				\fmfdot{v1}
				\fmf{scalar}{o1,v2}
				\fmf{dashes}{v2,v1}
				\fmf{fermion}{v1,v3,o3}
				\fmffreeze
				\fmf{photon}{v3,o2}
				%\fmfi{boson}{vpath (__i1,__v1) shifted (thick*(0.5,2))}
			\end{fmfgraph*}
		}
	\end{fmffile}
	\caption{\label{Vertices} The diagrams illustrating two allowed types of elementary processes providing a positron in the final state. Here $f$ denotes any fermion enabled by the symmetry group and kinematics and $\phi$ any enabled integer spin field.}
\end{figure}
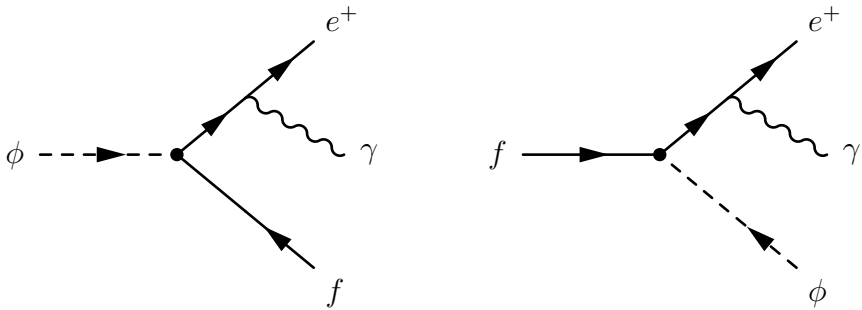

Now, we want to estimate the ``minimal'' model-independent initial spectra of prompt gamma radiation. %As we discussed above, the least possible flux of prompt radiation  is provided by the positrons, that eventually contribute to the AMS-02 signal. 
In these estimations we concentrate on the fact that the positron with a given energy was produced in an elementary process, which appears as some part of the DM annihilation cascade. For example, this process might be a $W^+$ or $Z$ decay or even the decay of some new positively charged massive particle. The set of possible vertices is, first of all, limited by Lorentz symmetry and renormalizability of the interaction. Thus, at the tree level, we are left with four point vertices (Fig. \ref{Vertices}).

\begin{figure}[h!]
	\centering
	\includegraphics[scale=0.40]{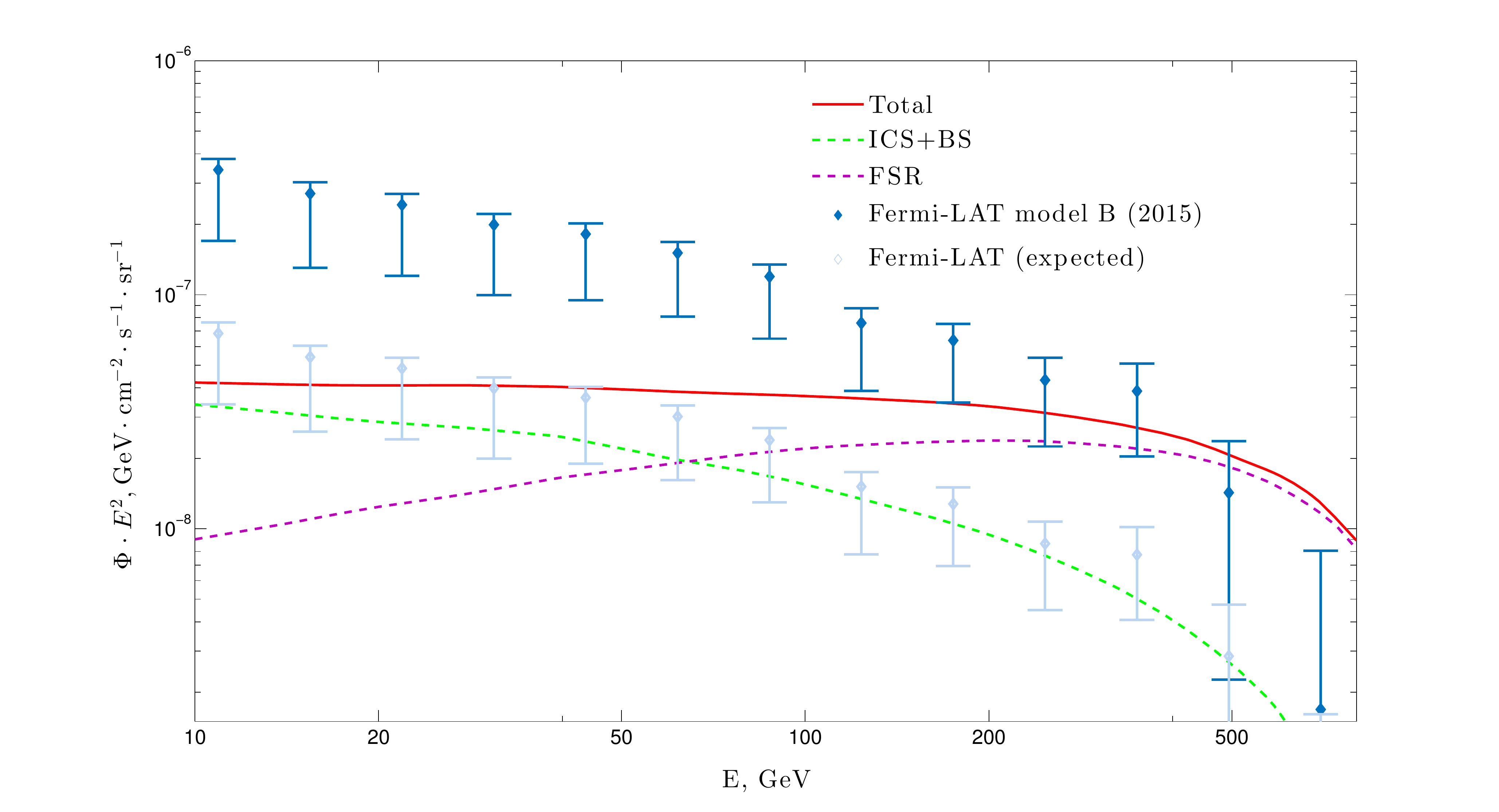}
	\caption{\label{gammaplot} The minimal gamma ray flux from DM annihilations compared to the contemporary (we take the most conservative model B, see \cite{Ackermann:2014usa})  and expected \cite{DiMauro:2016cbj} Fermi-LAT data on IGRB. Two major contributions are shown separately in different colors.  }
\end{figure}

Since we are interested in positrons and gamma rays with energies above 10 GeV, we expect that the energy spectrum of final state radiation from $e^+$ would only depend on the energy of the emitting positrons and that it can be calculated as

\begin{equation}
\label{mingamma}
\frac{dN}{dE} = \int\limits_E^{\rm 1~TeV} \,\phi_{\gamma}(E,E_0) f_e(E_0) \; dE_0,\\
\end{equation}
where $f_e (E_0)$ denotes the initial spectrum of positrons (see Fig. \ref{minpos}) and $\phi_{\gamma}(E,E_0)$ denotes the spectrum of photons produced by the positrons with energy $E_0$. It is given by \cite{Essig:2009jx}

\begin{equation} 
\phi_{\gamma}(E,E_0)=\frac{\alpha}{\pi E}  \left( 1 + \left( 1 - \frac{E}{E_0} \right)^2 \right) \left( \ln\left[ \left( \frac{2E_0}{m_e} \right)^2 \left( 1 - \frac{E}{E_0} \right) \right] -1 \right). \\%\nonumber
\end{equation}
Here we neglect the difference in the positron spectra \textit{before} and \textit{after} photon emission. The resulting minimal flux of gamma rays from DM annihilations is shown in Fig. \ref{gammaplot}.

As one can see, the minimal gamma-ray flux obtained satisfies the contemporary Fermi-LAT limit (except the last data point, which has a large error though), but, according to one of their recent papers \cite{DiMauro:2016cbj}, more than $80 \%$ of IGRB can be explained by unresolved astrophysical sources, such as active galactic nuclei. This new (expected) limit turns out to be much lower than the predicted minimal flux of gamma rays. \textit{Q.E.D.}

Also, one should take into account two facts, which make our result even stronger: a) we didn't take into consideration the extragalactic gamma-ray flux (since its estimations are model-dependent), which may be comparable to the Galactic one; b) practically, any DM model yields more prompt radiation than just the contribution from positrons. 

One of the key assumptions of our theorem is the conventional isotropic distribution of annihilating/decaying DM. As it was shown in one of our works~\cite{Belotsky:2016tja} one can circumvent this no-go theorem by assuming a non-isotropic positron source distribution, e.g. a dark matter disk. 

%\section{Conclusions}
%You might skip the section {\em conclusions} if you are running out of
%space.

\section*{Acknowledgments}
The author thanks all the co-authors of the original article for their valuable contributions: K.~Belotsky, R.~Budaev and A.~Kirillov. I thank J.-R.~Cudell for his ideas, inspiring this research. I am also grateful to A.~Bhattacharya, M.~Khlopov and D.~Wegman for fruitful discussions. This work is supported by a FRIA grant (F.N.R.S.). 

 \newpage

%\bibliographystyle{JHEP}
%\bibliography{References}

\providecommand{\href}[2]{#2}\begingroup\raggedright\endgroup

\end{document}